# "Electrical-modelling, design and simulation of cumulative radiation effects in semiconductor pixels detectors: prospects and limits"

**Nicolas T. Fourches** [a,*], **Remi Chipaux** [a]

[a] *CEA/IRFU, Saclay, 91191 GIF/YVETTE FRANCE*
*Corresponding author, E-mail*: `nicolas.fourches@cea.fr`

**ABSTRACT:** Silicon detectors have gained in popularity since silicon became a widely used micro/nanoelectronic semiconductor material. Silicon detectors are used in particle physics as well as imaging for pixel based detecting systems. Over the past twenty years a lot of experimental efforts have been focused on the effects of ionizing and non-ionizing radiation on silicon pixels. Some of this research was done in the framework of high luminosity particle physics experiments, along with radiation hardness studies of basic semiconductors devices. In its simplest form the semiconductor pixel detectors reduce to a PIN or PN structure partially or totally depleted, or in some MOS and APD (Avalanche PhotoDiode) structures. Bulk or surface defects affect considerably transport of free carriers. We propose guidelines for pixel design, which will be tested through a few pixel structures. This design method includes into the design the properties of defects. The electrical properties reduce to parameters, which can be introduced in a standard simulation code to make predictive simulations. We include an analytical model for defect annealing derived from isochronal annealing experiments. The proposed method can be used to study pixels detectors with different geometrical structures and made with different semiconducting materials. Its purpose is to provide an alternative to tedious and extensive radiation tests on final fabricated detectors. This is necessary for the long-term reliability of detectors together with their radiation tolerance. A general method for pixel design is introduced and we will show how it can be used for the design of alternate to silicon (germanium) pixels.

KEYWORDS: Avalanche-induced secondary effects, Charge transport and multiplication in solid media, Detector modelling and simulations II (electric fields, charge transport, multiplication and induction, pulse formation, electron emission, etc), Interaction of radiation with matter, Radiation-hard detectors, Solid state detectors, dE/dx detectors, Particle tracking detectors (Solid-state detectors), Particle tracking detectors, Pixelated detectors and associated VLSI electronic, Materials for solid-state detectors, pixel detector , avalanche photodiode, germanium, high luminosity, particle physics, non-ionizing radiation, ionizing radiation, deep defects, traps, radiation effects, radiation induced defects, radiation tests, defect annealing, simulation, technological simulation, physical simulation, PIN; multilayer structure.

**Contents**

**1. Charged particle detection with semiconductors**

An adequate pixel detector for future vertexing and tracking in particle physics experiments should meet the following prerequisites. First it should have a granularity not reached until now, down to the micron range. Second it should be radiation hard to levels not unknown until now, of more 500 Megarads (Si) in total ionizing dose and $10^{16}$ cm$^{-2}$ in 1MeV neutron equivalent fluence [1]. It should be also not sensitive to transient radiation effects, such as SEE (Single Event Effect) and avoid destructive latch-up. We shall focus on neutron-induced defects, which are the most detrimental for long-term S-C detectors.

**1.1 Requirements**

This imposes constraints on both the detecting material and the design of the readout electronic. Direct detection requires a semiconductor with a low energy threshold for electron-hole pair generation and with a low reverse dark leakage current, even after irradiation with massive particles such as neutrons. It sensitivity to radiation induced defects should be low. The operation mode of the detectors can be with internal amplification, high impact carrier generation coefficients should then be an advantage [2][3][4].

**1.2 Potential solutions and constraints**

Whereas high bandgap semiconductors allow fabrication of devices operating with low leakage current, in silicon or germanium high leakage currents may lead to increased noise and power dissipation. The dissipation on silicon detectors deduced from Moll measurements [5] result in a value close to 1.2 mW/cm$^2$ for a 50 reverse biased silicon detector with 50 µA leakage current [5] (irradiated @10$^{14}$ neutrons/cm$^2$, 1MeV equivalent). This shows that the main source of heat in pixel detectors is mainly due to the readout circuitry. That heat can be evacuated with a heat-conducting holder. The reduction of power dissipation can also be obtained in some cases by a simple scheme such as "switch on power during readout" like scheme if the detector is able to operate this way. This can be achieved in CCDs and CMOS sensors, with the exclusion of Hybrid Pixels that need a charge sensitive amplifier to operate.

## 2. Choices for the detecting material

### 2.1 Detecting material

Table I summarizes the mean ionization energies required to create an electron-hole pair in different semiconducting materials, and other physical quantities as well as their breakdown electric field, useful for particle detection evaluation.

**Table I:** characteristics of materials considered here for pixel detectors, some data are taken from [2], [6]. $E_{hn}$ is the average energy for electron-hole pair generation, LET is here the Linear Energy Transfer for a Minimum Ionizing Particle (MIP), the last column value is derived from [7] using the LET expressed in MeVg$^{-1}$cm$^2$ multiplied by the density of the material (in g.cm$^{-3}$). The maximum of electron-hole pair created per unit length can be computed by dividing the LET by $E_{hn}$. The density can be found in many basic references.

| Material | Si | Ge | CdTe | GaAs | SiC β | (diamond) C |
|---|---|---|---|---|---|---|
| Density in gcm$^{-3}$ | 2.33 | 5.33 | 5.85 | 5.32 | 3.21 | 3.5 |
| Bandgap | 1.1 eV | 0.67 eV | 1.44 eV (dir) | 1.4 eV (dir) | 2.3 eV | 5.47 eV |
| Breakdown field (MV/cm) | 0.3 | 0.1 | 0.4 | 0.4 | 2 | 20 |
| $E_{hn}$ | 3.6 eV | 2.98 eV | ~4.5 eV | ~4.5 eV | 8.8 eV | 12 eV |
| LET( MeV/cm) LET(MeVg$^{-1}$cm$^2$) | 3.6 1.6 | 7.5 1.4 | 7.3 1.25 | 7.5 1.4 | 5.5 1.7 | 6.3 1.8 |
| Number of electron-hole pairs generated | <~105 e-h / µm | <~ 250 e-h / µm | <~ 162 e-h / µm | <~167 e-h / µm | <~63 e-h / µm | <~52 e-h / µm |

The average energy for electron-hole pair creation for germanium is the lower value (estimated with 77 K operating detectors), although references in [6], suggest a value of 2.8 eV at room temperature. According to this table, simply taken into account the data from the literature, deducing an approximate LET for each material, and taken the average energy for electron-hole pair creation, the best material in terms of sensitivity should be germanium. For CdTe , GaAs, Diamond and SiC , the Nhn is lower than that of Germanium. We will next focus on this material.

**2.2 Energy deposition**

Using the data from Table I the mean number of free electron-hole pairs generated along a track of a MIP is of the order of: ~ 250 e-h / µm, for Ge and ~100 e-h / µm for Si (close to the 80 e-h/µm found for silicon). Early theoretical studies [8] computed an energy deposition MPV (Most Probable Value) in 10 µm silicon being the order of 1.857 keV with a FWHM of the order of 1.758 keV  (particle with charge +/-e  and βγ > 500).  Experimentally some results with 5 GeV electrons have been obtained with silicon CMOS sensors [9], which show that for small pixels < 25 µm x 25 µm, the detection efficiency reaches more than 98% with an Active Layer Thickness (ALT, detecting volume) of a few microns (=<10 µm). The output signal is statistically strong enough to obtain adequate detection efficiencies. The MPV is 388 e- corresponding to ~> 1.4 keV at least in energy deposition, in agreement with [8]. For a germanium based detecting media we have made GEANT4 simulations (Figure 1(left) and Figure 1(right)) for two thicknesses with 130 GeV pions. See Table II for result.

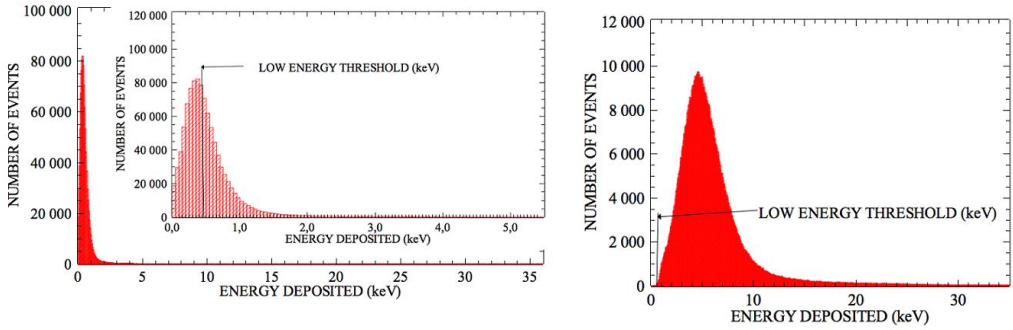

Figure 1: distribution of the deposited energy for 130 GeV pions in (left) a 1 micrometer germanium thick material, (right) 10 micrometer thick material: the most probable value is ~350 eV (left) and ~4.5 keV (right). The bin width is the same for the two figures (0.05 keV) (total number of events simulated with GEANT4 50000).  The threshold for detection efficiency is indicated (insert for 1µm thick Ge). Above 35 keV the number of events is negligeable. The average number of e-h pair generated in 10 µm Ge is 250 is ~2500 (250/µm) in agreement with Table I.

Table II: detection properties deduced from GEANT4 simulations, in germanium, a noise threshold of 15 electrons is a realistic assumption for pixel detectors [9] along with a signal to noise ratio of 10. This leads to a 150 electrons (0.45 keV in germanium) threshold. The detection efficiency is obtained by integration from the threshold to a 50 keV deposited limit.

| Thickness (micrometers) | Threshold (keV) | Detection efficiency (%) |
|---|---|---|
| 1 (Figure 1(left)) | 0.45 | 44 (estimation) |
| 10   (Figure 1 (right)) | 0.45 | 99 (estimation) |

This shows that the detecting material thickness (ALT) should be set close to 10 µm to remain optimal in terms of detection efficiency.

### 2.3 Germanium: signal formation

For new pixel detectors, a germanium device say 10μm x10μm in size (thickness 10 μm) is potentially convenient for charge particle detection progress has been made recently in the reduction of leakage current (see Table III for details).

Table III: summary of results of this work and literature [10]-[11], second row result from first row downscaling.

| Structure | Dark reverse current | Surface current magnitude | Expected signal magnitude (C) for a MIP |
|---|---|---|---|
| PIN (Ge) (300 K), thickness 0..3 cm Cylindric structure with BF3 implanted p+ contact and Li diffused n+ contact Al sputtered on the contact, high purity germanium, intrinsic at 300 K $n_i=10^{13}$ cm$^{-3}$ (a) | <1 +/- 0.5 mA/cm$^2$ (this work) -2 V reverse bias (measured), dependent on temperature  Capacitance@77K = 15 pF Net Doping @77K =$10^{10}$cm$^{-3}$ fromC(V) measurements at 77 K, near intrinsic region | n.a. @ 300 K, negligible w.r.t. volume current  @77 K < 1μA/cm (measured) | < 40 fC (250000 e) @ 0.1 cm active thickness computed from last column of Table I. LET/Ehn x active thickness |
| PIN (Ge): 10 μm x10 μm thickness 10 μm (b) | < tens of pA/μm$^2$ (deduced from first row) reduced by increased doping in the near intrinsic zone | n.a., deduced from (10) ~ 1 nA | < 0.4 fC (2500 e) ~ 6x10$^6$ e per second |
| Ge on Si (10) (c) | 1 nA (measured in (10)) | n.a. | n.available |
| With passivation (11) (d) | n.a. | ~1μA/cm(measured in (11)) | n.a. |

In [10] the dark leakage current could be reduced to approximately 1nA for a device close to (b). With appropriate passivation (using a GeO$_2$ layer) the surface current may be reduced to less than 1 μA/cm (where the unit length is the circumference of the structure) [11]. For squared 10 μm x 10μm structures this leads to an assumed leakage current of ~1nA lower than the signal (~20 ns duration I >> 20nA). This would also lead to an equivalent noise current lower than that of the readout with a signal to noise ratio of (S/N ~ 2500/15 e/e). Leakage current related problems may thus be controlled.

### 3. Simulation of irradiation induced defects

### 3.1 Recent advances

Radiation tolerance is a key point for inner tracker-vertex-detectors. As these are frequently implemented with hybrid architecture they can be studied independently

w.r.t. the electronic readout. A test and trial procedure has up to now been used for the development of present pixel detectors. However given the amount of experimental data obtained on silicon, full simulation of a silicon pixel as a predictive method may now be current. Radiation induced defects characteristics can be introduced in simulation program. Their electrical properties can be measured for detectors with DLTS and derived techniques [12][13][14][15]. According to [12] germanium the deepest hole trap found in neutron irradiated germanium, located 0.35 eV above the valence band, with an emission rate extrapolated from the Arrhenius plot of ~$10^6$ s$^{-1}$ at 300K. Most of the degradation by the deep levels should originate in the leakage current due to generation inside the depleted zone of the detector. Additionally, in silicon detectors, the charge collection efficiency versus neutron fluence can be fitted with a model parameterized with discrete energy levels in the bandgap, [13] related to a limited number of defects.

**3.2 Modelling the irradiated detector**

Future Upgraded High Energy Physics experiments require detector operation up to more than $10^{16}$ cm$^{-2}$ (1MeV equivalent neutrons) on a time scale which is 256x$10^6$ seconds ~10 years). With well-characterized detecting materials the physical characteristics of the radiation induced traps (capture cross sections, activation energy of the emission rate etc…) may be used for simulation. The limits are due to the standard spectroscopic techniques that do not detect easily some more complicated objects than point defects such as extended defects, which are on the opposite revealed by electron microscopy. We focus on the annealing behaviour of the defect where Nt is the defect concentration. The annealing function A(t) after irradiation is given by : $N_t(t) = A(t) = N_{t0} e^{-\alpha t}$ where $\alpha = \alpha_0 \exp\left(-\frac{E_a}{kT}\right)$ Ea being the activation energy for defect annealing, and $\alpha_0$ the kinetic coefficient at infinite temperature for a first order kinetic. In many cases the presence of clusters may screen the presence of deep defect levels [16]. From [12] [16] it can be shown that the capture rates for holes for identical defects were higher in neutron irradiated germanium than in electron-irradiated germanium. However the measured characteristics of secondary defects were sufficient to explain the degradation of the gamma ray detectors. For silicon Moll [5] showed that the characteristics of defects "= (microscopic levels)" present after irradiation and annealing measured by DLTS or TSC (Thermally Stimulated Current) were sufficient to parameterize a leakage current model. One step further is to take into the modelling scheme the annealing behaviour of the defects, when these parameters are experimentally known. In n-type silicon the high temperature annealing stage reaches more than 600 K (leaving aside the lower temperature stages [17] below 330 K), much higher than in germanium. Most defects introduced at room temperature in Ge disappear in the 100°C-200°C annealing stage although some reverse annealing appears around 100°C (due to vacancy cluster break up [16]). In [12] this was only noticeable for the HT2/LT2 level ($E_v$+0.16 eV) although it slightly anneals in the 250K-300K range (Figure11 in [12]). The relatively high annealing activation energy (HT2) is due in part by the reverse annealing thus allowing this value to be taken as a realistic value for the global HT2 annealing process. For a given fluence the total number of vacancies available by cluster breakup is finite (introduction rate < 35 cm$^{-1}$) setting a higher limit to the HT2 concentration. In lightly doped neutron irradiated germanium a screening effect due to the presence of p-type zones in n-type material appears. This makes the electron-traps difficult to detect. The amorphous nature of the damaged regions was found questionable in early studies, the density of defects, though high [18], indicates that no transition to an amorphous phase occurs at these low fluences. Table IV give a summary for neutron-irradiated germanium.

**Table IV**: short review of deep defects found in high purity neutron irradiated germanium using transient capacitance techniques, capture cross sections obtained from capture kinetic (from [12]). Comparison is made with proton and electron irradiation, with the H190 level related to the divacancy [14] fits with HT2 [12] and H1 [15] as being also related to the divacancy. The origin of the HT1, HT3 and HT4 was discussed in [12]. The introduction rate is the ratio of the defect concentration to the neutron fluence. The concentration versus fluence fit is linear in the fluence range investigated. The defect annealing temperature is the temperature value for which the concentration of the defect reduces to half of its value before annealing (Isochronal anneal). Using the isochronal anneal data of [12] for each defect we have in this work computed the kinetic coefficient for each temperature calculated the corresponding activation energies for $\alpha$ using a first order annealing model, $\alpha_0$ being also deduced from this data.

| Electron traps | Hole traps From [12] | Energy level (eV) | Capture cross section (cm$^2$) | Introduction rate (cm$^{-1}$) | Annealing temperature (°C) | Annealing activation energy (eV) |
|---|---|---|---|---|---|---|
| | HT1 | 0.35 | >10$^{-14}$ | 0.27 | ~160 °C | 0.46-0.58 |
| | HT2 (V$_2$) | 0.17-0.18 | >10$^{-14}$ | 0.45 | ~175 °C | 0.78-0.81 |
| | HT3 | 0.12 | >10$^{-14}$ | 0.35 | ~150 °C | 0.44-0.60 |
| | HT4 | 0.25 | >10$^{-14}$ | 0.16 | ~145 °C | 0.34-0.33 |
| Screened (77-200 K) | | | or < 10$^{-17}$ | | | |
| | H190 (V$_2$) [14] | 0.185 | | | ~115 °C | |
| | H140 (I$_2$) | 0.14 | | | | |
| | H1(V$_2$) | 0.16 | | | | |

Huhtinen [19] addressed the problem of defects generation using numerical simulations where basic interactions (atomic collisions in the solid) are taken as the starting point. These ab initio techniques lead to a description of defects introduced in silicon by different hadronic particles. However full simulation requires the knowledge of the electrical properties of final defects generated by hadrons in the material and this implies measurements on irradiated samples. We take a simplified approach experimental-measurement approach, where the introduction rate for defects is taken constant in a wide fluence range. This is reasonable if we consider that the impurity concentration is high w.r.t. that of primary defects, leading to an unchanged generation rate of defect complexes. This should be valid for neutral impurities in high concentration in Ge (~10$^{14}$ cm$^{-2}$). The low annealing temperature of point defects should have an influence on the final defect concentration. We take into consideration the deepest defect level in the bandgap, with a defect concentration equal to $N_t$. If R is the introduction rate, and φ the fluence rate: $φ=39\times10^6$ cm$^{-2}$ s$^{-1}$.

$$N_t(t) = Rφt * A(t) \quad (1)$$

Where A(t) is the annealing function and * stands for the convolution product. This gives: $N_t(t) = Rφα^{-1}[1 - \exp(-αt)] (2)$, if we consider defect annealing with first order kinetic $E_a$ is the activation energy. If this equation is taken into consideration at room temperature the cumulated concentration is much lower than expected with the former equation. A step forward would be to accurately model the detailed annealing processes (including reverse annealing). The limitation is the lack of extensive experimental data for germanium that is necessary for accurate simulation. At 300 K, for the deepest hole trapped found in neutron irradiated germanium HT1 [12], the concentration after $10^{16}$ cm$^{-2}$ (1MeV equivalent neutrons), with R=1cm$^{-1}$, can be given by equ.(2) with α=84x10$^{-9}$ s$^{-1}$, is <~4x10$^{14}$ cm$^{-3}$, with first order annealing model and ~10$^{16}$ cm$^{-3}$ with no annealing. This proves that the end defect concentration may vary by more than one order of magnitude. This also means that irradiation tests made at a high fluence rate (total fluence in a few hours) are not a valid qualification procedure.

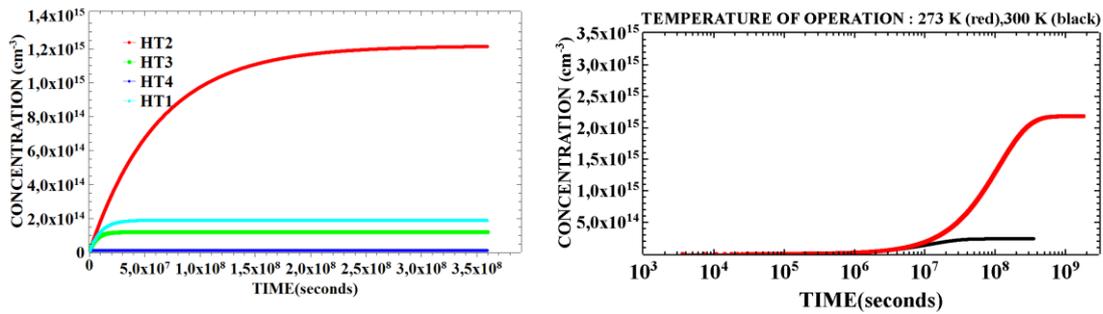

Figure 2: Plot of the concentration of deep defects versus time using an introduction/annealing model described in the text, the reference temperature is 300 K. On the left concentration versus time for the for hole traps (Table IV). On the right( HT1 ) red: the simulation temperature is 273 K. black: the simulation temperature is 300 K. The fluence rate 3.8 x10$^7$ cm$^{-2}$s$^{-1}$, corresponds to a total dose of $10^{16}$ cm$^{-2}$ for 10 years of operation (300 days a year of continuous operation) without taking the "no beam" time into consideration. For simplicity the same introduction rate is taken for each defect (0.5cm$^{-1}$).

The simulations establish that the temperature of operation has a significant influence on the final defect concentration (Figure 2). A factor of 10 can be estimated in favour of 300 K operation with respect to 273 K operation for a fluence rate of 3.8x10$^7$ cm$^{-2}$s$^{-1}$. However a conservative view is to consider that a thermal treatment up to 150 °C would be sufficient to reduce significantly carrier trapping. This would be compatible with a silicon based readout for which a periodic thermal cycling at 150°C of the electronics would then be possible. On this basis germanium could then be an attractive alternative material to silicon for pixel detectors.

### 3.3 Simulation of avalanche structures

Device simulation can be carried out with standard simulation codes available, PISCES for electrical simulation and SUPREM4 for the process simulations (commercial software suites from Silvaco can be used) which are necessary to assess the manufacturing feasibility of the device. The full simulation of a silicon pixel has been performed in [20] with a homemade software suite. The critical electric field for breakdown is lower in germanium than in silicon [Table I]. However for a simple germanium PIN structure (rectangular with a 5µm active thickness: 5µmx1µmx5µm) the breakdown occurs at high reverse voltage (50-80 Volts) making such a structure unsuitable for signal multiplication even if the thickness is reduced. The effect of deep defects [Table IV] is marked at very high concentration ($10^{16}$ $n_{eq}/cm^2$) leading to high leakage currents (a simulation at $10^{16}$ $n_{eq}/cm^2$ result in a generation current of more than 10nA) making the device unusable in non-avalanche mode. The design of a multi-doped-layer structure is thus possible with a much reduced breakdown voltage [Figure 3] and an internal gain of 12.

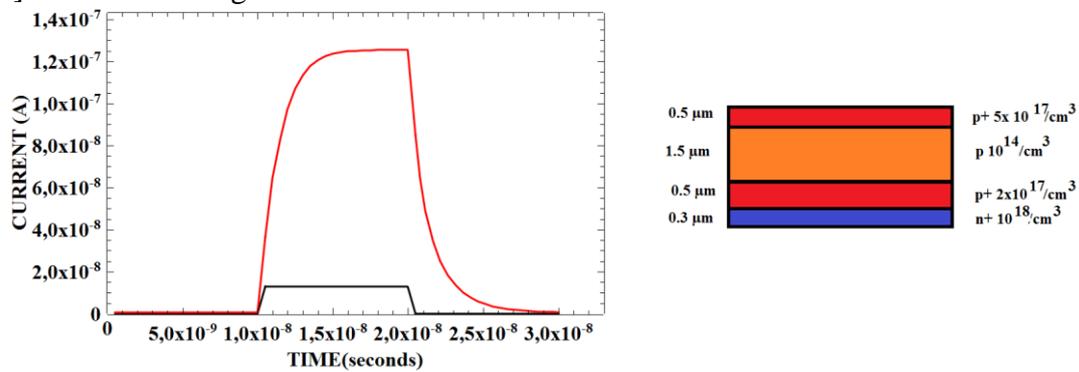

Figure 3: plot of the response of the APD versus time for a Minimum Ionizing Particle Track in avalanche mode. The reverse voltage was set to ~9 Volts with dimensions being 1µmx1µmx3 µm (3µm active thickness). The structure was a p+ip+n+ as shown (left) with the amplifying zone being the middle p+ zone.

## 4. Summary and conclusions

We have made an overview of specific material choices on the simple basis of carrier generation/energy deposition properties. We have shown that an alternative to silicon exists. Despite being considered as a second in terms of processing and leakage current control, we have shown that germanium with its low indirect gap can be still considered as a good semiconducting medium for charged particle detection. The effect of radiation-induced defects due to hadrons can be reduced in germanium compared with other semiconductors because the majority of the displacement-induced defects anneal out at reasonably low temperatures. A small internal amplification detector can be designed on the basis of an APD structure with detecting and multiplying zone. The multiplication factor can be greater than 10 for these devices. The problem encountered in designing such structures is the reduction of the critical voltage for avalanche operation. This is necessary for integration into standard silicon processes. The effect of noise increased by the presence of deep levels on the fake events generation should then be studied in the view of reliable operation with high counting rates.

**Acknowledgments**

The authors acknowledge the contribution of the internship student Yuchao Yang for the PISCES simulations. F. Eozenou provided assistance for Ge etching chemistry and P.F. Honoré provided a valuable technical support.

**References**

[1] ATLAS, *Letter Of Intent, Phase II, upgrade CERN-2012-022, LHCC-1-023*, December 2012

[2] A.S. Kyuregyan and S.N. Yurkov, *Room-temperature avalanche breakdown voltages of p-n junctions made of Si, Ge, SiC, GaAs, GaP, and lnP*, Sov. Phys. of Sem. 23(10), 1126-1131, (1989)

[3] B. Ziaja, R.A. London, & J. Hajd, *Ionization by impact electrons in solids: Electron mean free path fitted over a wide energy range,* J. Appl. Phys., 99(3), 033514, (2006) http://dx.doi.org/10.1063/1.2161821

[4] C. L. Anderson' and C. R. Crowell, *Threshold Energies for Electron-Hole Pair Production by Impact Ionization in Semiconductors,* Phys. Rev. B., Vol. 5(6) 2267 (1972) http://dx.doi.org/10.1103/PhysRevB.5.2267

[5] M. Moll, *Radiation damage In Silicon Particle Detectors*, Doctoral Dissertation, University of Hamburg, Fachbereich Physics, 1999

[6] C. A. Klein, *Bandgap Dependence and Related Features of Radiation Ionization Energies in Semiconductors,* J. Appl. Phys. 39, 2029 (1968) http://dx.doi.org/10.1063/1.1656484

[7] C. Amsler et al. Review of Particle Physics, *Passage of particle through matter*, Physics Letters B, Vol. 667, Issues 1-5, 268 (2008) http://dx.doi.org/10.1016/j.physletb.2008.07.018

[8] H. Bichsel, *Straggling in thin silicon detectors,* Reviews of Modern Physics, 60(3), 663, (1988) http://dx.doi.org/10.1103/RevModPhys.60.663

[9] N. Fourches (Spokeperson) et al., *Performance of a Fast Programmable Active Pixel Sensor Chip Designed for Charged Particle Detection,,* Contributed Talk to the Nuclear Science Symposium, October 23-29, 2005, Porto Rico, IEEE NSS Record, N4-7, page 93 - 97 , Vol.1 (2005) http://dx.doi.org/10.1109/NSSMIC.2005.1596214

[10] S.B Samavedam, M.T. Currie,T.A. Langdo, & E.A. Fitzgerald, *High-quality germanium photodiodes integrated on silicon substrates using optimized relaxed graded buffers*, Appl. Phys. Lett., 73(15), 2125-2127(1998) http://dx.doi.org/10.1063/1.122399

[11] Mitsuru Takenaka, Kiyohito Morii, Masakazu Sugiyama,Yoshiaki Nakano,and Shinichi Takagi, *Dark current reduction of Ge photodetector by GeO2 surface passivation and gas-phase doping,* Optics Express, Vol. 20, Issue 8, pp. 8718-8725 (2012) http://dx.doi.org/10.1364/OE.20.008718

[12] N.Fourches and G. Walter, J.C. Bourgoin , *Neutron induced defects in high purity germanium* , J. Appl. Phys. 69 (4) 2033 (1991) http://dx.doi.org/10.1063/1.348728 and Nicolas Fourches, Doctoral Thesis, Strasbourg 1989.

[13] NT. Fourches, *Device simulation of Monolithic Active Pixel Sensors: Radiation damage effects*, IEEE Transactions On Nuclear Science, Vol. 56, No.6, Pages 3743-3751, December 2009 http://dx.doi.org/10.1109/TNS.2009.2031540

[14] M. Christian Petersen, A. Nylandsted Larsen, and A. Mesli, *Divacancy defects in germanium studied using deep-level transient spectroscopy*, Phys. Rev. B 82, 075203 – Published 6 August 2010, http://dx.doi.org/10.1103/PhysRevB.82.075203

[15] F. Poulin, and J. C. Bourgoin, *Minority Carrier Traps in Electron Irradiated n-Type Germanium*, Recent Developments in Condensed Matter Physics. Springer US, p83-88 (1981)

[16] N. Fourches, *High defect density regions in neutron irradiated high purity germanium: characteristics and formation mechanisms,* J. Appl. Phys. 77 (8) 3684 (1995) http://dx.doi.org/10.1063/1.358607

[17] K. Gill, G. Hall, and B. MacEvoy, *Bulk damage effects in irradiated silicon detectors due to clustered divacancies*, J. of Appl. Phys. 82, 126 (1997); http://dx.doi.org/10.1063/1.365790

[18] A.R. Peaker, et al., *Vacancy Clusters in Germanium*, Vol. 131, pp. 125-130 (2008), doi http://dx.doi.org/10.4028/www.scientific.net/SSP.131-133.125

[19] M. Huhtinen, *Simulation of non-ionizing energy loss and defect formation in Silicon,* Nucl. Instr. and Meth. In Phys. Res. A491, 194-214 (2002), http://dx.doi.org/10.1016/S0168-9002(02)01227-5

[20] M. Swartz, *A detailed simulation of the CMS Pixel Sensor*, CMS Note 2002/027, John Hopkins University Baltimore Maryland